\documentclass[a4paper,11pt]{article}
\usepackage{pos}

\usepackage{amsmath,amssymb,physics,bm}
\usepackage{graphicx}
\usepackage{float}
\usepackage{hyperref}

\newcommand{\be}{\begin{equation}}
\newcommand{\ee}{\end{equation}}
\newcommand{\mA}{\mathcal{A}}
\newcommand{\mO}{\mathcal{O}}
\newcommand{\mD}{\mathcal{D}}

\DeclareMathOperator{\im}{Im}
\DeclareMathOperator{\sgn}{sgn}

\title{Constrained Symplectic Quantization: Disclosing the Deterministic Framework Behind Quantum Field Theory}
\ShortTitle{Constrained Symplectic Quantization: Quantum Field Theory}

\author*[a,b]{Francesco Scardino}
\author[c,d]{Martina Giachello}
\author[e]{Giacomo Gradenigo}

\affiliation[a]{Physics Department, INFN Roma1, Piazzale A. Moro 2, Roma, I-00185, Italy}
\affiliation[b]{Physics Department, Sapienza University, Piazzale A. Moro 2, Roma, I-00185, Italy}
\affiliation[c]{Gran Sasso Science Institute, Viale F. Crispi 7, 67100 L'Aquila, Italy}
\affiliation[d]{INFN-Laboratori Nazionali del Gran Sasso, Via G. Acitelli 22, 67100 Assergi (AQ), Italy}
\affiliation[e]{Physics and Astronomy Department "Galileo Galilei", Universit\`a di Padova, Via Marzolo 8, 35131 Padova, Italy}

\emailAdd{francesco.scardino@uniroma1.it}
\emailAdd{martina.giachello@gssi.it}
\emailAdd{giacomo.gradenigo@unipd.it}

\abstract{
Symplectic quantization is a functional approach to quantum field theory that allows sampling of quantum fluctuations directly in Minkowski space time by means of a Hamiltonian dynamics in an intrinsic time $\tau$ which samples a microcanonical ensemble, in close analogy with the standard microcanonical approach to lattice field theory. In this contribution we present constrained symplectic quantization for relativistic quantum field theory \cite{Scardino:2026pqr}, generalizing from the quantum mechanical case. The method is based on the analytic continuation of fields and action from $\mathbb{R}$ to $\mathbb{C}$ and on constraints that select stable intrinsic time trajectories and that simultaneously define convergent integration cycles for the microcanonical partition function. In the continuum limit we recover the Feynman generating functional with the correct real time prescription. We test the construction for a free scalar field in $1+1$ dimensions on a periodic lattice by measuring real time two point functions and by verifying Dyson Schwinger identities with the correct contact term.
}

\FullConference{The 42nd International Symposium on Lattice Field Theory (LATTICE2025)\\
2--8 November 2025\\
Tata Institute of Fundamental Research, Mumbai, India\\}

\begin{document}
\maketitle

\section{Introduction}
Symplectic quantization is a new functional formulation of quantum field theory based on an extended variable space in which quantum fluctuations are parametrized by an additional intrinsic time $\tau$ \cite{Gradenigo_1,Gradenigo_2,Gradenigo_3}. In the case of a relativistic scalar quantum field theory (QFT), the operator $\hat\varphi(x)$ is replaced by a $\tau$ dependent field $\varphi(x,\tau)$ together with a conjugate momentum field $\pi(x,\tau)$, and the intrinsic time evolution is governed by a set of generalized Hamilton equations. Assuming a mild ergodicity hypothesis, one identifies long $\tau$ averages with averages in a microcanonical ensemble at fixed generalized action, in close analogy with microcanonical strategies in lattice field theory \cite{deAlfaro:1983zz,Strominger:1982xu,Iwazaki:1984kx,Callaway:1983,Duane:1987de}.\par

A purely real realization of symplectic quantization leads, in the large number of degrees of freedom limit, to a canonical weight proportional to $\exp\{S/\hbar\}$ rather than the oscillatory Feynman weight $\exp\{iS/\hbar\}$ and it becomes ill defined in the free theory limit \cite{Giachello_JHEP,Giachello_lattice2024,Giachello_trani}.The constrained symplectic quantization (CSQ) framework introduced in CSQ I solves the issue by analytically continuing the fields and the action from $\mathbb{R}$ to $\mathbb{C}$ while keeping the generalized Hamiltonian real and imposing constraints on the intrinsic time flow \cite{Scardino:2026pqr,Giachello_CSQ1}. This provides at once a stable deterministic dynamics and a direct correspondence with the Feynman path integral measure.\par

Real time quantum field theory poses a well known obstacle to numerical approaches due to the oscillatory nature of the path integral weight~\cite{Gattringer:2016kco,Alexandru:2020wrj}. The constrained symplectic framework attacks the problem by replacing the sampling of a complex weight with a deterministic intrinsic time flow whose invariant measure reproduces the same generating functional. The logic is close in spirit to stochastic and microcanonical strategies~\cite{Creutz:1980gp,Duane:1987de,Parisi:1980ys}, but the basic evolution is Hamiltonian and it preserves a generalized action exactly in the continuum limit.\par

Minkowski signature requires more than a change of variables. One needs a choice of integration cycle in a complexified field space that renders the generating functional well defined and that implements the Feynman prescription. The logic of CSQ is closely related to modern contour-deformation strategies for the sign problem~\cite{Alexandru:2020wrj,Cristoforetti:2012su,Scorzato:2015qts,Behtash:2015loa,Witten:2010cx}: one complexifies the field space and restricts the dynamics to an appropriate stable manifold. A conceptual distinction is that in Lefschetz-thimble methods the constraint is imposed on the {\it integration cycle} of the path integral, while in CSQ the constraint is imposed directly on the {\it microscopic Hamiltonian flow} in intrinsic time, which generates the effective sampling measure. The constrained manifold provides bounded evolution in intrinsic time and it selects the same holomorphic cycle that appears in the functional integral representation. \par

In this contribution we focus on the microcanonical generating functional, on its equivalence with the Feynman functional integral, and on the free scalar field dynamics and tests presented in our talk. We recall the microcanonical generating functional, its large number of modes limit, and the emergence of the Feynman weight. We then focus on the intrinsic time identities that follow from stationarity, with particular emphasis on the Dyson Schwinger relations used as a non trivial check of the simulations for the free scalar theory.
%
%
%
\section{Microcanonical formulation and complexification}
As described in ~\cite{Giachello_CSQ1} for the case of quantum mechanics, we promote the intrinsic time fields and momenta to complex variables
\be
\varphi(x,\tau)=\varphi_R(x,\tau)+i\,\varphi_I(x,\tau),\qquad
\pi(x,\tau)=\pi_R(x,\tau)+i\,\pi_I(x,\tau),
\ee
and we treat $\varphi$ and its complex conjugate as independent integration variables along one dimensional contours in the complex plane.
The generalized Hamiltonian of constrained symplectic quantization is
\be
\mathbb{H}_{\rm SQ}[\pi,\bar\pi,\varphi,\bar\varphi]
=
\int d^{d+1}x\;\bar\pi(x,\tau)\pi(x,\tau)
+2\,\im S[\varphi,\bar\varphi],
\qquad
\im S[\varphi,\bar\varphi]=\frac{S[\varphi]-\bar S[\bar\varphi]}{2i}.
\label{eq:hsq}
\ee
Assuming ergodicity of the constrained Hamiltonian flow, long intrinsic time averages of a holomorphic observable $\mO[\varphi]$ are identified with the corresponding microcanonical functional integral average
\be
\langle O[\varphi]\rangle_\tau=\lim_{\tau\rightarrow\infty}\frac{1}{\tau}\int_{0}^{\tau} d\tau'\,\mO[\varphi(x,\tau')]
=
\int \mD\bar\varphi\,\mD\varphi\,\mD\bar\pi\,\mD\pi\;
\rho_{\text{micro}}[\varphi,\bar{\varphi},\pi,\bar{\pi}]\,\mO[\varphi]=\langle O[\varphi]\rangle.
\label{eq:erg}
\ee
The microcanonical density and partition function at fixed $\mA$ are
\be
\rho_{\text{micro}}=\frac{1}{\Omega[\mA,0]}\,
\delta\,\Big(\mA-\mathbb{H}_{\rm SQ}\Big),
\qquad
\Omega[\mA,0]=
\int_{\boldsymbol{\Gamma}(x)} \mathcal{D}\bar{\pi}\,\mathcal{D}\pi\,\mathcal{D}\bar{\varphi}\,\mathcal{D}\varphi\;
\delta\,\Big(\mA - \mathbb{H}_{\rm SQ} \Big).
\label{eq:micro}
\ee
The analytic continuation from $\mathbb{R}$ to $\mathbb{C}$ does not imply a doubling of integration degrees of freedom since each field is still integrated along a one dimensional contour $\boldsymbol{\Gamma}(x)$.
The constraints select stable intrinsic time trajectories and fix the relevant integration cycles in the complexified field space.

\section{From the microcanonical functional to the Feynman weight}

In this section we generalize to relativistic quantum field
theory the equivalence proven in the $0+1$-dimensional case in
Ref.~\cite{Giachello_CSQ1}.  Our goal
is to show that, once the generalized Hamiltonian of constrained
symplectic quantization is fixed to $\mA=\hbar M$, the
microcanonical generating functional constructed from the conservation
of $\mathbb{H}_{\rm SQ}$ reproduces, in the continuum limit, the
standard Feynman generating functional of a scalar field in Minkowski
space-time. More precisely, introducing an external source $J(x)$, we
prove that
\begin{align}
  \lim_{M\rightarrow\infty}\,
  \frac{1}{\Omega[\hbar M,0]}\,
  \frac{\delta^n\Omega[\hbar M,J]}{\delta J(x_{1})\ldots\delta J(x_{n})}\Bigg|_{J=0}
  \;=\;
  \big\langle \varphi(x_1)\ldots \varphi(x_n)\big\rangle
  \;=\;
  \frac{1}{\mathcal{Z}[\hbar,0]}\,
  \frac{\delta^n\mathcal{Z}[\hbar,J]}{\delta J(x_{1})\ldots\delta J(x_{n})}\Bigg|_{J=0}\,,
  \label{eq:equivalence-sq-fey-qft}
\end{align}
where $\mathcal{Z}[\hbar,J]$ denotes the Feynman path integral of the
holomorphic theory defined by the analytically continued Minkowskian
action $S[\varphi]$.

We start by defining the microcanonical generating functional of CSQ in
$d+1$ dimensions as
\begin{align}
\Omega[\mA,J] =
\frac{1}{\Omega[\mA,0]}
\int_{\boldsymbol{\Gamma}(x)} \mathcal{D}\bar{\pi}\,\mathcal{D}\pi\,\mathcal{D}\bar{\varphi}\,\mathcal{D}\varphi\;
\delta\,\Big(\mA - \mathbb{H}_{\rm SQ}[\pi,\bar{\pi},\varphi,\bar{\varphi}] + i\, J\!\cdot\!\varphi\Big)\,,
\label{eq:micro-part-func-qft}
\end{align}
where $x\equiv (x_0,\mathbf{x})$ is a Minkowskian space-time point and
\begin{align}
  J\!\cdot\!\varphi \equiv \int d^{d+1}x\, J(x)\,\varphi(x)\,,\qquad
  \bar{\pi}\!\cdot\!\pi \equiv \int d^{d+1}x\,\bar{\pi}(x)\,\pi(x)\,.
\end{align}
The generalized Hamiltonian is the one introduced in
the previous section,
\begin{align}
  \mathbb{H}_{\rm SQ}[\pi,\bar{\pi},\varphi,\bar{\varphi}] =
  \bar{\pi}\!\cdot\!\pi + 2\,\Im S[\varphi,\bar{\varphi}]\,,\qquad
  \Im S[\varphi,\bar{\varphi}] = \frac{S[\varphi]-\bar{S}[\bar{\varphi}]}{2i}\,.
  \label{eq:hsq-qft}
\end{align}
As in Ref.~\cite{Giachello_CSQ1}, $\boldsymbol{\Gamma}(x)$ denotes
the set of one-dimensional integration contours in the complex plane
parametrized by the space-time point $x$, one contour for each of the
fields to be integrated, $\varphi(x)$, $\bar{\varphi}(x)$, $\pi(x)$ and
$\bar{\pi}(x)$. We write
\begin{align}
\boldsymbol{\Gamma}(x) = \Gamma_\varphi(x)\cup \Gamma_{\bar{\varphi}}(x)\cup \Gamma_\pi(x)\cup \Gamma_{\bar{\pi}}(x)\,,
\label{eq:contours-qft}
\end{align}
so that, for instance,
\begin{align}\label{eq:feyn}
\Gamma_\varphi(x)=\left\lbrace \gamma_\varphi(x)\in\mathbb{C}: x\in\mathcal{V}_{d+1}\right\rbrace,
\end{align}
with $\mathcal{V}_{d+1}$ a finite space-time volume. Let us stress again
a crucial point: the analytic continuation from $\mathbb{R}$ to
$\mathbb{C}$ does not imply any doubling of integration degrees of
freedom. The integration domain for each field at each space-time point
remains a one-dimensional path and so the only difference is that this path is
embedded in $\mathbb{C}$ rather than being the real line. The additional
variables in Eq.~\eqref{eq:micro-part-func-qft} are the intrinsic-time
momenta $\pi,\bar{\pi}$, which will be integrated out explicitly.

We integrate out the intrinsic time momenta by representing the delta function as a Fourier integral over a Lagrange multiplier $\lambda$
\be
\Omega[\mA,J]
\propto
\int \mathcal{D}\bar{\pi}\,\mathcal{D}\pi\,
\mathcal{D}\bar{\varphi}\,\mathcal{D}\varphi\,
\frac{d\lambda}{2\pi}\;
\exp\!\Big[
-i\lambda\,\bar{\pi}\!\cdot\!\pi
+i\lambda\big(\mA-2\,\im S[\varphi,\bar{\varphi}]+ i\,J\!\cdot\!\varphi\big)
\Big],
\label{eq:lambda}
\ee
with $\bar{\pi}\!\cdot\!\pi=\int d^{d+1}x\,\bar{\pi}(x)\pi(x)$.
The Gaussian integral over $\pi$ factorizes over the degrees of freedom and yields a power law factor $\lambda^{-M}$ on a lattice with $M$ modes, as in~\cite{Giachello_CSQ1}.
Performing the $\lambda$ integral one obtains
\be
\Omega[\mA,J]
\propto
\int \mathcal{D}\bar{\varphi}\,\mathcal{D}\varphi\;
\Big(\mA-2\,\im S[\varphi,\bar{\varphi}]+ i\,J\!\cdot\!\varphi\Big)^{M-1}.
\label{eq:Omega-power}
\ee
Setting $\mA=\hbar M$ and extracting an overall factor, one can write
\be
\Big(\hbar M-2\,\im S[\varphi,\bar{\varphi}]+ i\,J\!\cdot\!\varphi\Big)^{M-1}
=
(\hbar M)^{M-1}
\left(
1-\frac{2}{\hbar M}\im S[\varphi,\bar{\varphi}]
+\frac{i}{\hbar M}J\!\cdot\!\varphi
\right)^{M-1}.
\label{eq:power-expand}
\ee
Using $\lim_{M\to\infty}\left(1+\frac{x}{M}\right)^M=e^x$ one obtains in the large-$M$, continuum limit the following
\be
\Omega[\hbar M,J]
\cong
\frac{1}{\Omega_\infty[\hbar,0]}
\int \mathcal{D}\bar{\varphi}\,\mathcal{D}\varphi\;
\exp\!\left[
-\frac{2}{\hbar}\im S[\varphi,\bar{\varphi}]
+\frac{i}{\hbar}J\!\cdot\!\varphi
\right]\,,
\label{eq:Omega-asympt}
\ee
which holds true for the renormalized theory. By expanding the definition of $\Im S$ we get
\begin{align}
\Omega[\hbar M,J]
&\cong
\frac{1}{\Omega_\infty[\hbar,0]}
\left[
\int_{\Gamma_\varphi}\mathcal{D}\varphi\;
\exp\!\left(\frac{i}{\hbar}S[\varphi]+\frac{i}{\hbar}\int d^{d+1}x\,J(x)\varphi(x)\right)
\right]
\left[
\int_{\Gamma'_{\bar{\varphi}}}\mathcal{D}\bar{\varphi}\;
e^{-\frac{i}{\hbar}\bar{S}[\bar{\varphi}]}
\right].
\label{eq:Omega-asympt-2-qft}
\end{align}
Finally, the anti-holomorphic functional integral factorizes both in the
numerator and in the denominator of the normalized definition
Eq.~\eqref{eq:micro-part-func-qft}, hence it cancels out. We thus obtain
the continuum-limit generating functional
\begin{align}
\lim_{M\to\infty}\Omega[\hbar M,J]
=
\Omega_\infty[\hbar,J]
=
\frac{
\int_{\Gamma_\varphi}\mathcal{D}\varphi\;
\exp\!\left(\frac{i}{\hbar}S[\varphi]+\frac{i}{\hbar}\int d^{d+1}x\,J(x)\varphi(x)\right)
}{
\int_{\Gamma_\varphi}\mathcal{D}\varphi\;
\exp\!\left(\frac{i}{\hbar}S[\varphi]\right)
}\,.
\label{eq:Omega-to-Z-qft-final}
\end{align}
By choosing $\Gamma_\varphi$ to implement the $i\epsilon$ prescription
(i.e. the standard Feynman integration cycle), Eq.~\eqref{eq:Omega-to-Z-qft-final}
is precisely the normalized Feynman generating functional
$\mathcal{Z}[\hbar,J]/\mathcal{Z}[\hbar,0]$. This proves
Eq.~\eqref{eq:equivalence-sq-fey-qft}, namely that the microcanonical
correlators generated by constrained symplectic quantization coincide,
in the continuum limit, with the Minkowskian correlators of the Feynman
path integral.

The argument just presented is crucial because it places on firm grounds
the correspondence between the microcanonical partition function
underlying CSQ and the Feynman path integral in $d+1$ dimensions. At the
same time, it is formal in the same sense as in Ref.~\cite{Giachello_CSQ1}:
it does not yet specify how the integration cycles $\Gamma_\varphi$ and
$\Gamma'_{\bar{\varphi}}$ should be selected constructively in the
microcanonical formulation. In constrained symplectic quantization, this
selection is realized dynamically by the underlying constrained
Hamiltonian flow in intrinsic time. We discuss this point in detail in
the following sections.

\section{Intrinsic time flow and Dyson Schwinger identities}
Hamilton equations generated by Eq. \eqref{eq:hsq} imply
\be
\partial_\tau \varphi=\frac{\delta\mathbb{H}_{\rm SQ}}{\delta \pi}=\bar\pi,
\qquad
\partial_\tau \pi=-\frac{\delta\mathbb{H}_{\rm SQ}}{\delta \varphi}=i\,\frac{\delta S[\varphi]}{\delta\varphi},
\label{eq:ham}
\ee
together with the conjugate equations.
In the stationary regime, the intrinsic time total derivative of a holomorphic observable has vanishing average.
The Dyson Schwinger relations follow from the same stationarity statement used to define intrinsic time averages. Let $\mO[\varphi]$ be a holomorphic functional of the field. We consider the intrinsic time derivative of the product $\pi(x,\tau)\mO[\varphi]$ and we average it over a long intrinsic time interval. By recalling the time average definition in Eq. \eqref{eq:erg}, in the stationary regime the boundary term vanishes and we obtain
\be
0=\left\langle \partial_\tau\big(\pi(x,\tau)\mO[\varphi]\big)\right\rangle
=
\left\langle \partial_\tau \pi(x,\tau)\,\mO[\varphi]\right\rangle
+
\left\langle \pi(x,\tau)\,\partial_\tau \mO[\varphi]\right\rangle.
\label{eq:ds-start}
\ee
Using Eq. \eqref{eq:ham} we rewrite the first term as
\be
\left\langle \partial_\tau \pi(x,\tau)\,\mO[\varphi]\right\rangle
=
i\left\langle \frac{\delta S}{\delta\varphi(x)}\,\mO[\varphi]\right\rangle.
\label{eq:ds-first}
\ee
The second term is evaluated by the chain rule. Since $\mO$ is holomorphic we have
\be
\partial_\tau \mO[\varphi]=\int d^{d+1}z\;\frac{\delta \mO[\varphi]}{\delta\varphi(z)}\,\partial_\tau\varphi(z,\tau),
\qquad
\partial_\tau\varphi(z,\tau)=\bar\pi(z,\tau).
\label{eq:ds-chain}
\ee
The microcanonical measure is invariant under the constrained Hamiltonian flow and the momentum variables are Gaussian once the delta constraint is resolved. This implies that the mixed contraction between $\pi$ and $\bar\pi$ produces a local contact term. The resulting identity is the standard Dyson Schwinger relation at $\mA=\hbar M$ in the continuum limit  ($M   \to\infty$):
\be
\left\langle \frac{\delta S}{\delta\varphi(x)}\,\mO[\varphi]\right\rangle
=
i\,\hbar\left\langle \frac{\delta \mO[\varphi]}{\delta\varphi(x)}\right\rangle.
\label{eq:ds-general}
\ee
Choosing $\mO[\varphi]=\varphi(y)$ gives the two point relation
\be
\big\langle \frac{\delta S}{\delta\varphi(x)}\,\varphi(y)\big\rangle
=
i\,\hbar\,\delta^{(d+1)}(x-y),
\label{eq:ds}
\ee
which is the identity used in the numerical study. For higher insertions one obtains the full hierarchy, for instance
\be
\big\langle \frac{\delta S}{\delta\varphi(x)}\,\varphi(y)\varphi(z)\big\rangle
=
i\,\hbar\Big[
\delta^{(d+1)}(x-y)\,\langle \varphi(z)\rangle
+
\delta^{(d+1)}(x-z)\,\langle \varphi(y)\rangle
\Big].
\label{eq:ds2}
\ee
For a theory with vanishing one point function this reduces to contact terms only and it becomes a sharp probe of the normalization fixed by $\hbar$.

In the free scalar case one has
\be
\frac{\delta S}{\delta\varphi(x)}=
\big(\partial_{x_0}^2-\nabla^2+m^2\big)\varphi(x),
\label{eq:free-dS}
\ee
so Eq. \eqref{eq:ds} implies the Green function equation
\be
\big(\partial_{x_0}^2-\nabla^2+m^2\big)\,G(x-y)=i\,\hbar\,\delta^{(d+1)}(x-y).
\label{eq:free-green}
\ee
On a periodic lattice this becomes a discrete identity for the measured correlator. It is convenient to test it in momentum space where each mode satisfies an algebraic relation. Using the lattice dispersion relation one expects $\big(\hat{k}_0^2-\hat{k}_1^2-m^2\big)\,G(k)=i\,\hbar$, up to the standard zero mode subtleties on a finite volume. In our implementation the same constraints that stabilize the intrinsic time dynamics also fix the integration cycle, so the Dyson Schwinger relations provide a consistency check of both the stability and the Feynman prescription at the level of observables.
\section{Free scalar field and constraints}
We consider the free scalar theory with Minkowski action
\be
S[\varphi]=\frac{1}{2}\int d^{d+1}x\;
\varphi(x)\,\big(\partial_{x_0}^2-\nabla^2+m^2\big)\,\varphi(x),
\label{eq:freeS}
\ee
analytically continued to a holomorphic functional of the complex field.
Combining the holomorphic and anti holomorphic Hamilton equations one obtains the mixed equation of motion
\be
-i\,\frac{d^2}{d\tau^2}\bar{\varphi}(x,\tau)
=
\big(\partial_{x_0}^2-\nabla^2+m^2\big)\varphi(x,\tau),
\label{eq:eomx}
\ee
together with its complex conjugate.
Generic solutions are unbounded in $\tau$ unless one selects a stable manifold in the complexified phase space.
This selection parallels the choice of convergent integration contours in the microcanonical functional integral.

For periodic boundary conditions on a finite box we expand in Fourier modes
\be
\varphi(x,\tau)=\frac{1}{\sqrt{V_{d+1}}}\sum_{k}\,e^{i k\cdot x}\,\varphi(k,\tau),
\qquad
\omega^2(k)=k_0^2-\mathbf{k}^2-m^2,
\label{eq:fourier}
\ee
so that each mode satisfies
\be
\frac{d^2}{d\tau^2}\,\bar{\varphi}(k,\tau) + i\,\omega^2(k)\,\varphi(k,\tau)=0.
\label{eq:mode}
\ee
Writing $\varphi(k,\tau)=\varphi_R(k,\tau)+i\,\varphi_I(k,\tau)$, the flow is stable if real and imaginary components are tied together by a linear constraint.
Mode by mode we impose
\be
\varphi_I(k,\tau)=-\sgn[\omega^2(k)]\,\varphi_R(k,\tau),
\qquad
\pi_I(k,\tau)=-\sgn[\omega^2(k)]\,\pi_R(k,\tau),
\label{eq:stable}
\ee
which reduces the dynamics to bounded oscillatory motion with frequency $|\omega(k)|$.
In mode space, Eq. \eqref{eq:stable} corresponds to a rotation of the integration contour by an angle $\pi/4$ with a sign fixed by $\omega^2(k)$, as in the harmonic oscillator case \cite{Giachello_CSQ1}.
It is useful to introduce rotated variables $\varphi_\pm(k,\tau)=e^{\mp i\pi/4}\varphi(k,\tau)$.
On the constrained manifold one component is projected out and the remaining one evolves as a standard harmonic mode in $\tau$.
The sign choice in Eq. \eqref{eq:stable} implements the correct $i\epsilon$ prescription for each mode and defines the integration cycle $\Gamma_\varphi$ appearing in Eq. \eqref{eq:feyn}.

\section{Numerical algorithm and results in $1+1$ dimensions}
\label{sec:num_1p1}
\begin{figure}[H]
\centering
\includegraphics[width=0.80\linewidth]{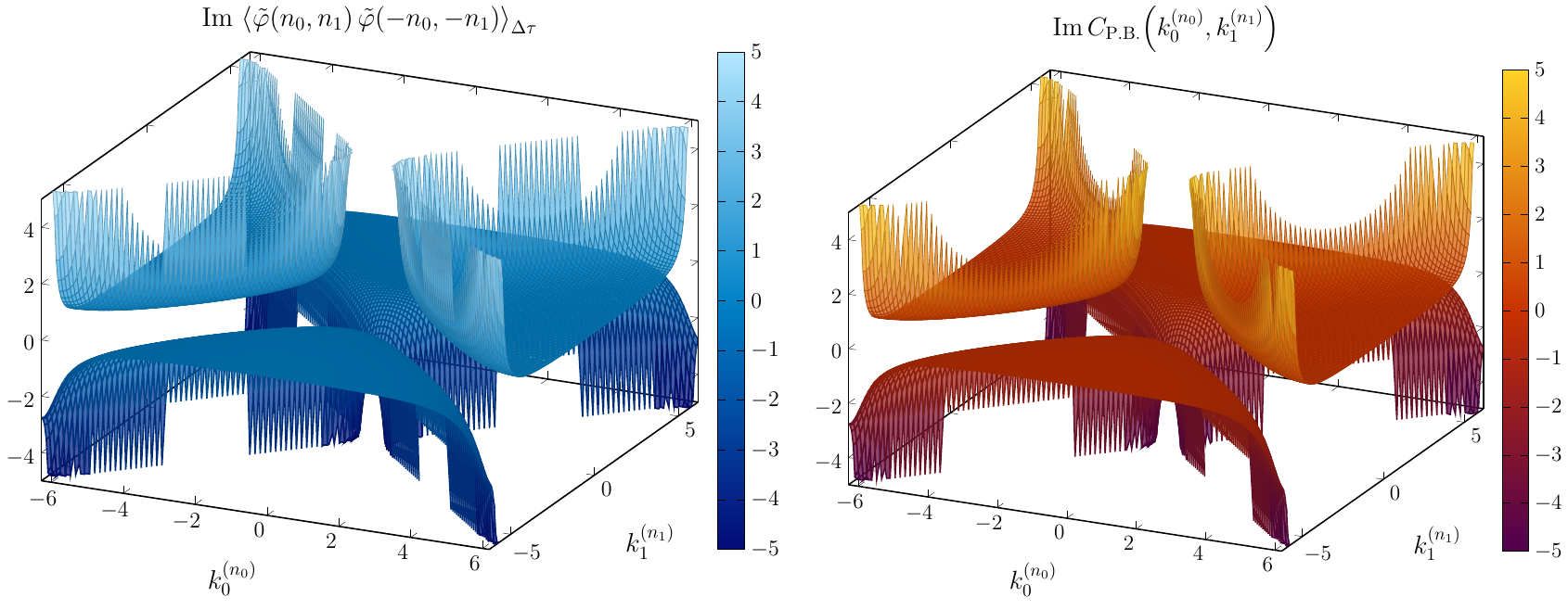}
\caption{Momentum-space propagator with periodic boundary conditions. Numerical estimate of $\Im\langle\tilde\varphi(n_0,n_1)\tilde\varphi(-n_0,-n_1)\rangle_{\Delta\tau}$ compared with the analytic lattice prediction $\Im\,\widetilde C_{\mathrm{P.B.}}(k_0^{(n_0)},k_1^{(n_1)})=1/(\hat k_0^2-\hat k_1^2-m^2)$, with $\hat k_\mu=\frac{2}{a}\sin(k_\mu a/2)$. The lightest singularity at $n_1=0$ ($k_1=0$) identifies the mass gap. Simulation parameters: mass $m=0.6$, lattice size $a=0.5$ on a $N_0\times N_1=65536$ square lattice.}
\label{fig:prop-mom}
\end{figure}

We simulate the constrained intrinsic--time dynamics of the free scalar field in Eq.~\eqref{eq:eomx} on a periodic $N_0\times N_1$ lattice in $1+1$ Minkowski dimensions, with spacing $a$ and extents $T=N_0a$, $L=N_1a$. The complex field $\varphi(\ell_0,\ell_1;\tau)$ and its conjugate momentum $\pi(\ell_0,\ell_1;\tau)$ are evolved in intrinsic time with a symplectic integrator, monitoring that the constraint of Eq.~\eqref{eq:stable} is implemented for each mode. The first observable is the generalized Hamiltonian $\mathbb H_\tau$ of Eq.~\eqref{eq:hsq}, monitored along the trajectory as a global control of integration errors. This check certifies that the intrinsic-time flow remains stable and confined to the intended constrained manifold.
\begin{figure}[H]
\centering
\includegraphics[width=0.5\linewidth]{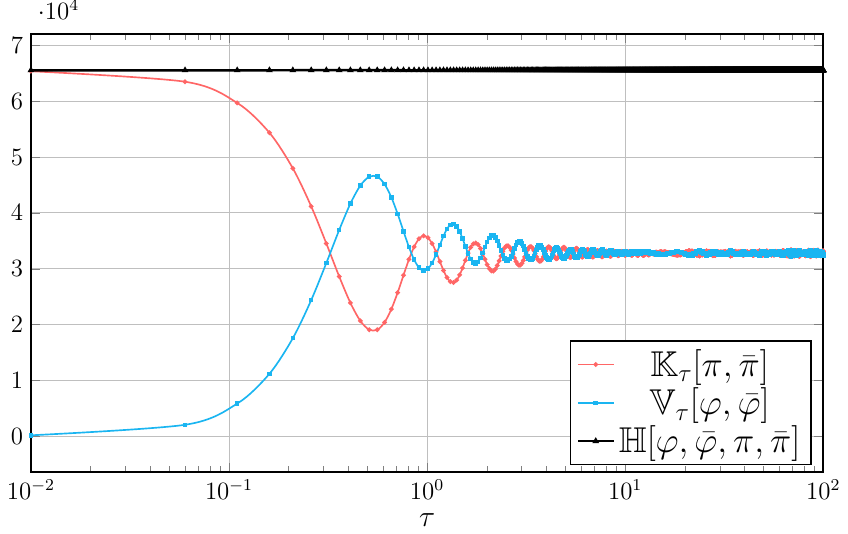}
\caption{Generalized action conservation along the constrained intrinsic--time evolution. The generalized Hamiltonian $\mathbb H_\tau$ is conserved up to step-size errors, providing a global diagnostic of the reversible symplectic integrator under repeated Fourier-space projections.}
\label{fig:energy}
\end{figure}
Two-point functions are computed as intrinsic--time averages in the stationary regime,
\begin{equation}
G(x-y)=\lim_{\Delta\tau\to\infty}\frac{1}{\Delta\tau}\int_{\tau_0}^{\tau_0+\Delta\tau}\! d\tau\;
\varphi(x,\tau)\,\varphi(y,\tau).
\label{eq:Gdef}
\end{equation}
On the periodic lattice it is convenient to work in discrete Fourier space. With $x_0^{(\ell_0)}=\ell_0 a$, $x_1^{(\ell_1)}=\ell_1 a$ and discrete momenta
\begin{equation}
k_0^{(n_0)}=\frac{2\pi n_0}{T},\qquad n_0=-\frac{N_0}{2},\dots,\frac{N_0}{2}-1,
\qquad
k_1^{(n_1)}=\frac{2\pi n_1}{L},\qquad n_1=-\frac{N_1}{2},\dots,\frac{N_1}{2}-1,
\end{equation}
we define the lattice Fourier modes $\tilde\varphi(n_0,n_1)$. For the free theory, the analytic \emph{lattice} momentum-space correlator is diagonal and reads
\begin{equation}
\Big\langle \tilde\varphi(n_0,n_1)\,\tilde\varphi(-n_0,-n_1)\Big\rangle_{\mathrm{P.B.}} =
\widetilde C_{\mathrm{P.B.}}\!\left(k_0^{(n_0)},k_1^{(n_1)}\right), \qquad \widetilde C_{\mathrm{P.B.}}(k_0,k_1) = \frac{i}{\hat k_0^2-\hat k_1^2-m^2},
\label{eq:lat-prop-in-section}
\end{equation}
with the standard coordinates $\hat k_\mu=\frac{2}{a}\sin\!\left(\frac{k_\mu^{(n_\mu)} a}{2}\right)$. We then compare numerical correlators with Eq.~\eqref{eq:lat-prop-in-section} finding perfect agreement, as visible in Fig.~\ref{fig:prop-mom}.
On a fully periodic lattice the coordinate-space correlator $C(\ell_0,\ell_1)=\langle\varphi(\ell_0,\ell_1)\varphi(0,0)\rangle$ contains competing temporal and spatial images. To obtain one-dimensional signals with a direct spectral interpretation we consider the two marginalized correlators
\begin{equation}
C_T(\ell_0)\equiv \frac{1}{N_1}\sum_{\ell_1=0}^{N_1-1}C(\ell_0,\ell_1),
\qquad
C_L(\ell_1)\equiv \frac{1}{N_0}\sum_{\ell_0=0}^{N_0-1}C(\ell_0,\ell_1).
\end{equation}
Using discrete orthogonality, these sums act as projectors in momentum space: the spatial sum enforces $n_1=0$ (zero spatial momentum, $k_1=0$), while the temporal sum enforces $n_0=0$ (zero frequency, $k_0=0$). Equivalently,
\begin{equation}
C_T(\ell_0)=\frac{1}{N_0}\sum_{n_0} e^{ik_0^{(n_0)}x_0^{(\ell_0)}}\,
\Big\langle \tilde\varphi(n_0,0)\tilde\varphi(-n_0,0)\Big\rangle,
\qquad
C_L(\ell_1)=\frac{1}{N_1}\sum_{n_1} e^{ik_1^{(n_1)}x_1^{(\ell_1)}}\,
\Big\langle \tilde\varphi(0,n_1)\tilde\varphi(0,-n_1)\Big\rangle.
\end{equation}
As a result, $C_T$ isolates the $k_1=0$ slice of the lattice propagator and displays a clean oscillation governed by the mass gap, whereas $C_L$ isolates the $k_0=0$ slice and controls the spatial correlation length.
\begingroup
\setlength{\intextsep}{2pt}      
\setlength{\abovecaptionskip}{0pt}
\setlength{\belowcaptionskip}{0pt}
\begin{figure}[H]
\centering
\includegraphics[width=0.48\linewidth]{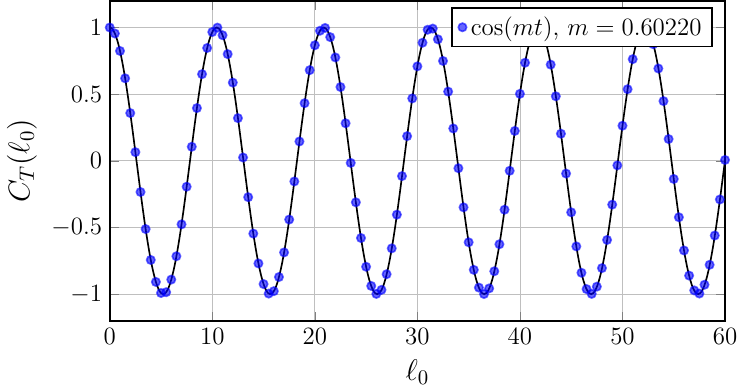}
\includegraphics[width=0.45\linewidth]{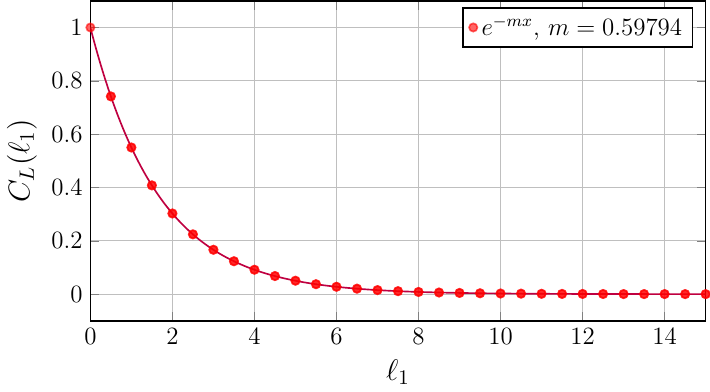}
\caption{Marginalized correlators on the periodic lattice.
\textbf{Left:} spatial marginalization $C_T(\ell_0)=\frac{1}{N_1}\sum_{\ell_1}C(\ell_0,\ell_1)$ projects onto $n_1=0$ ($k_1=0$) and isolates the temporal oscillation set by the mass gap. \textbf{Right:} temporal marginalization $C_L(\ell_1)=\frac{1}{N_0}\sum_{\ell_0}C(\ell_0,\ell_1)$ projects onto $n_0=0$ ($k_0=0$) and governs the spatial correlation length. Simulation parameters: mass $m=0.6$, lattice size $a=0.5$ on a $N_0\times N_1=65536$ square lattice.}
\label{fig:marg}
\end{figure}
\endgroup
Finally, we test the Dyson--Schwinger identity of Eq.~\eqref{eq:ds} by measuring the mixed correlator $\langle \delta S/\delta\varphi(\ell)\,\varphi(\ell_i)\rangle$. On the lattice the continuum delta distribution is represented as $\delta^{(2)}(x-x_i)\to \delta_{\ell,\ell_i}/a^2$, so the prediction is a purely local contact term. The presence of this peak provides a stringent validation of the stationary intrinsic--time measure and of the discretized equations of motion.
\begingroup
\setlength{\intextsep}{2pt}      
\setlength{\abovecaptionskip}{0pt}
\setlength{\belowcaptionskip}{0pt}
\begin{figure}[H]
\centering
\includegraphics[width=0.75\linewidth]{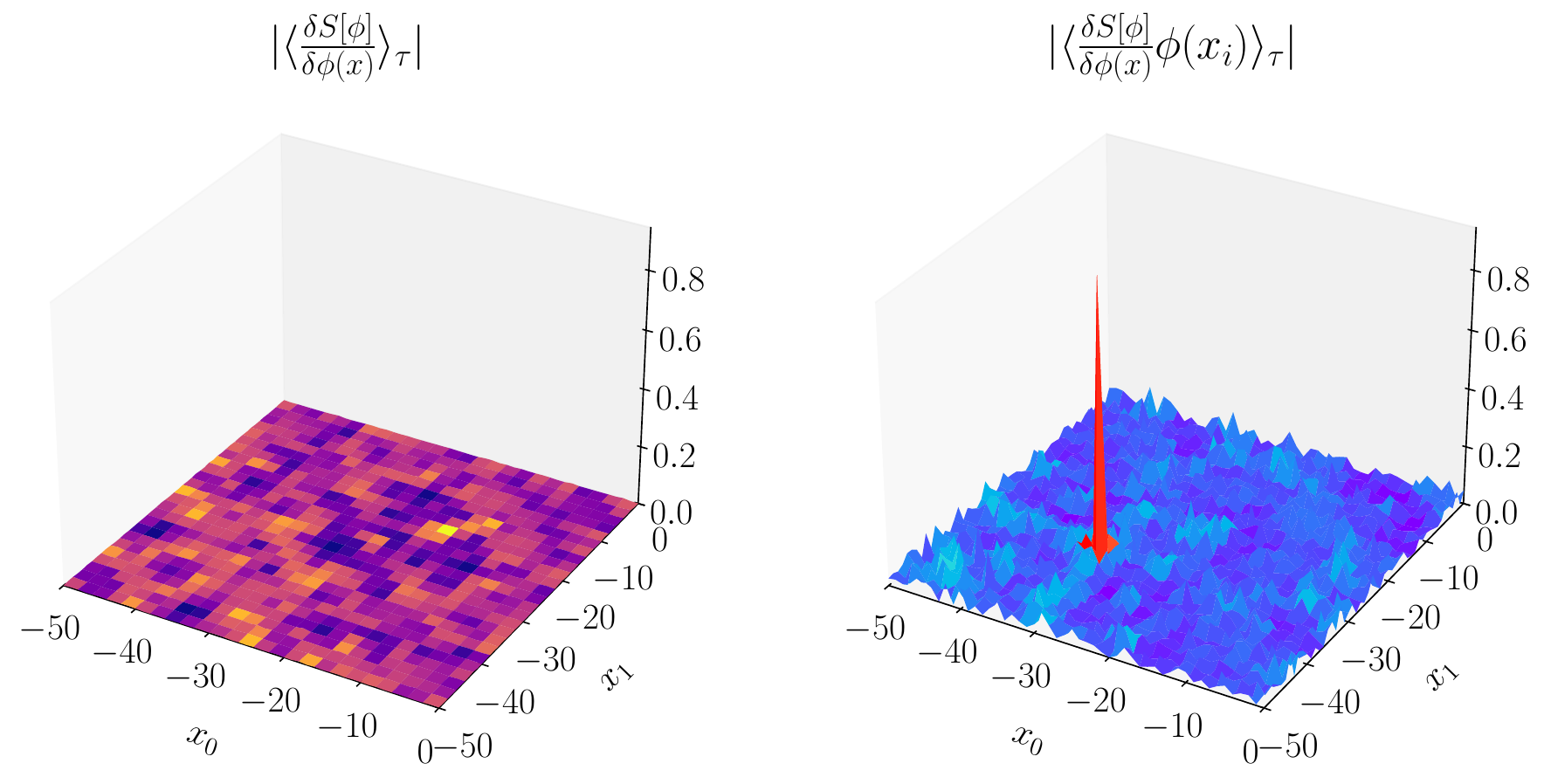}
\caption{Dyson--Schwinger numerical verification with periodic boundary conditions.
We measure $\langle \delta S/\delta\varphi(\ell)\,\varphi(\ell_i)\rangle$ and resolve the localized lattice contact term predicted by Eq.~\eqref{eq:ds2}.}
\label{fig:ds}
\end{figure}
\endgroup
\section{Conclusions and outlook}
Constrained symplectic quantization provides a stable deterministic dynamics in an intrinsic time variable and reproduces the Feynman weight through a suitable complexification and a constrained Hamiltonian flow.
For a free scalar field in $1+1$ dimensions we computed real time two point functions and verified Dyson-Schwinger identities including the contact term with periodic boundary conditions.
Interacting systems are currently under investigation and the main challenge is to identify stable manifolds in the complexified field space that remain compatible with the contour deformation underlying the equivalence with the path integral.

\begingroup
\small
\renewcommand{\baselinestretch}{0.98}\selectfont
\bibliographystyle{JHEP}
\bibliography{references}
\endgroup

\end{document}